\documentclass[sigconf,preprint]{acmart}
\AtBeginDocument{%
  }

\setcopyright{acmlicensed}
\copyrightyear{2026}
\acmYear{2026}
\acmDOI{XXXXXXX.XXXXXXX}
\acmConference[Conference acronym 'XX]{Make sure to enter the correct
  conference title from your rights confirmation email}{June 03--05,
  2018}{Woodstock, NY}

\acmISBN{978-1-4503-XXXX-X/2018/06}

\usepackage{amsmath,amsfonts,amssymb}
\usepackage{amsthm}
\usepackage{algorithm}
\usepackage{algorithmic}
\usepackage{graphicx}
\usepackage{booktabs}
\usepackage{multirow}
\usepackage{colortbl}
\usepackage{array}
\usepackage{balance}
\usepackage{natbib}
\usepackage{url}
\usepackage{hyperref}
\usepackage[table]{xcolor}

\definecolor{tableheader}{RGB}{46,134,171}
\definecolor{tablerowalt}{RGB}{245,248,250}
\definecolor{bestresult}{RGB}{46,134,171}
\definecolor{lightblue}{RGB}{232,240,234}

\begin{document}

\title{ADS-POI: Agentic Spatiotemporal State Decomposition for Next Point-of-Interest Recommendation}


\author{Zhenyu Yu}
\affiliation{%
  \institution{Fudan University}
  \country{China}
}

\author{Chunlei Meng}
\affiliation{%
  \institution{Fudan University}
  \country{China}
}

\author{Yangchen Zeng}
\affiliation{%
  \institution{Southeast University}
  \country{China}
}

\author{Mohd Yamani Idna Idris}
\affiliation{%
  \institution{University of Malaya}
  \country{Malaysia}
}

\author{Shuigeng Zhou}
\affiliation{%
  \institution{Fudan University}
  \country{China}
}

\renewcommand{\shortauthors}{Yu et al.}

\begin{abstract}
Next point-of-interest (POI) recommendation requires modeling user mobility as a spatiotemporal sequence, where different behavioral factors may evolve at different temporal and spatial scales. Most existing methods compress a user’s history into a single latent representation, which tends to entangle heterogeneous signals such as routine mobility patterns, short-term intent, and temporal regularities. This entanglement limits the flexibility of state evolution and reduces the model’s ability to adapt to diverse decision contexts. We propose ADS-POI, a spatiotemporal state decomposition framework for next POI recommendation. ADS-POI represents a user with multiple parallel-evolving latent sub-states, each governed by its own spatiotemporal transition dynamics.
These sub-states are selectively aggregated through a context-conditioned mechanism to form the decision state used for prediction. This design enables different behavioral components to evolve at different rates while remaining coordinated under the current spatiotemporal context. Extensive experiments on three real-world benchmark datasets from Foursquare and Gowalla demonstrate that ADS-POI consistently outperforms strong state-of-the-art baselines under a full-ranking evaluation protocol. The results show that decomposing user behavior into multiple spatiotemporally-aware states leads to more effective and robust next POI recommendation. Our code is available at https://github.com/YuZhenyuLindy/ADS-POI.git.
\end{abstract}


\begin{teaserfigure}
  \includegraphics[width=\textwidth]{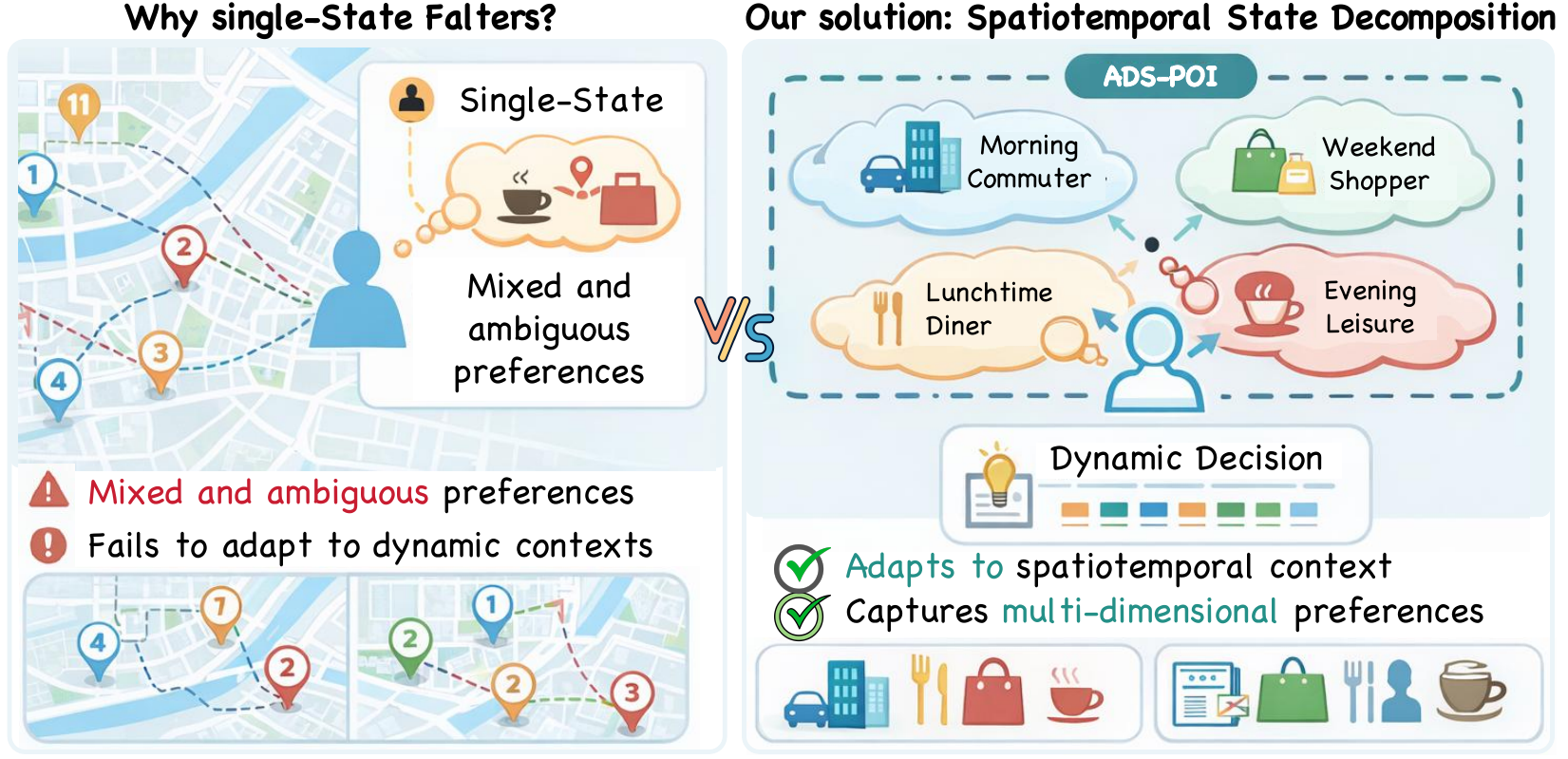}
  \caption{Motivation. Single-state user modeling mixes heterogeneous mobility preferences, while spatiotemporal state decomposition represents user behavior via multiple decision states and adaptively aggregates them under dynamic contexts.}
  \label{fig:teaser}
\end{teaserfigure}

\received{20 February 2007}
\received[revised]{12 March 2009}
\received[accepted]{5 June 2009}


\maketitle

\section{Introduction}

The rapid growth of location-based social networks (LBSNs) and mobile sensing technologies has produced massive volumes of spatiotemporal mobility data.
Leveraging such data, next point-of-interest (POI) recommendation aims to predict where a user will visit next based on historical check-in sequences and contextual signals~\cite{yang2022getnext,luo2021stan}.
Accurate POI recommendation underpins many real-world applications, including personalized navigation, urban analytics, and location-based services, and plays a critical role in supporting real-time, context-aware decision making in urban environments.

POI recommendation differs fundamentally from traditional item recommendation because human movement is constrained by \emph{space} and \emph{time}.
Users cannot arbitrarily travel to faraway locations, and their decisions are strongly shaped by temporal context (e.g., time-of-day and day-of-week) as well as geographic feasibility.
These spatiotemporal constraints impose structural regularities on mobility and require models to jointly capture spatial reachability, temporal periodicity, and personalized preference, turning next POI prediction into a challenging spatiotemporal sequential modeling problem that goes beyond static preference estimation.

Most existing POI models summarize a user’s historical trajectory into a single latent representation~\cite{liu2016strnn,sun2020lstpm}.
While effective, this single-state paradigm inevitably entangles heterogeneous signals: long-term preferences, habitual routines, spatial constraints, and short-term exploratory intent are compressed into one vector.
In practice, which factor dominates varies substantially across contexts and time, and a single representation has limited capacity to selectively surface the right signal at the right moment.
This entanglement often leads to brittle predictions, especially when the same user exhibits markedly different mobility patterns across time periods, activity types, or situational contexts.

\emph{A key observation motivating this work is that these behavioral factors evolve at different rates.}
Commuting between home and workplace is typically stable over long horizons; dining and leisure preferences may shift gradually; exploratory behavior can change abruptly from one visit to the next.
However, single-state models implicitly impose a largely uniform transition dynamic over all signals, forcing stable and fast-changing behaviors to share the same evolution path.
This mismatch makes it difficult to simultaneously track long-term routines and rapidly varying intent—an intrinsic property of real-world mobility—and limits a model’s ability to adapt under dynamic spatiotemporal conditions.
What we need is a representation that can model multiple behavioral processes in parallel, each with its own spatiotemporal dynamics, while still producing a coherent and discriminative prediction.

Based on this observation, we propose \textbf{ADS-POI}, an \emph{Agentic Spatiotemporal State Decomposition} framework for next POI recommendation.
Instead of compressing a trajectory into a single state, ADS-POI represents each user with multiple parallel latent sub-states, each specializing in a distinct behavioral dimension and capturing a particular aspect of mobility.
These sub-states evolve independently through state-specific spatiotemporal transition mechanisms, enabling the model to capture stable routines and rapidly changing preferences simultaneously.
At prediction time, ADS-POI performs \emph{context-adaptive aggregation}: it dynamically combines sub-states according to the current spatiotemporal context, emphasizing the most relevant behavioral signals for the current decision.
Here, “agentic” reflects an interpretation of the prediction process as coordinating multiple latent decision states; ADS-POI is trained purely in a supervised manner and does not rely on reinforcement learning or policy optimization.

We summarize our main \textbf{contributions} as follows:
\begin{itemize}
    \item \textbf{Spatiotemporal State Decomposition:} We propose a multi-state representation for next POI recommendation that decomposes user behavior into parallel latent sub-states, alleviating behavioral entanglement inherent in single-state models.
    \item \textbf{Heterogeneous Spatiotemporal Dynamics:} We design state-specific transition mechanisms so different behavioral dimensions can evolve at appropriate temporal and spatial rates, capturing the intrinsic multi-rate dynamics of human mobility.
    \item \textbf{Context-Adaptive Aggregation:} We introduce a context-adaptive aggregation strategy that dynamically combines multiple sub-states for next POI prediction, and demonstrate through extensive experiments on three real-world benchmark datasets that ADS-POI consistently outperforms state-of-the-art methods while yielding interpretable behavioral patterns.
\end{itemize}

\section{Related Work}

\subsection{POI Recommendation}

\textbf{Early Sequential Models.}
Next point-of-interest (POI) recommendation has been extensively studied in location-based services.
Early approaches extended item recommendation techniques with sequential signals, typically modeling check-ins as a user--POI interaction sequence with transition patterns.
FPMC~\cite{rendle2010fpmc} combined matrix factorization with first-order Markov chains to capture short-range dependencies, showing that recent visits provide strong cues for next-location prediction.
PRME~\cite{feng2015prme} embedded POIs into a metric space to model personalized transitions and distance-aware effects, offering a simple yet effective way to encode geographic proximity into sequential recommendation.
These early methods established a key insight that sequential dynamics and spatial constraints are both indispensable in mobility prediction, but they rely on relatively limited transition structures and shallow representations.
More importantly, they typically assume that user mobility can be governed by a single transition pattern, which restricts their ability to model heterogeneous behavioral modes within the same trajectory.

\textbf{Recurrent Neural Models.}
With deep learning, recurrent architectures became a dominant paradigm for sequential POI recommendation by learning nonlinear trajectory dynamics from raw sequences.
ST-RNN~\cite{liu2016strnn} incorporated spatial and temporal contexts via distance- and time-specific transition matrices, demonstrating that mobility transitions vary significantly across spatiotemporal conditions.
LSTPM~\cite{sun2020lstpm} separated long-term preference and short-term interest using nonlocal operations and geo-dilated recurrent units, highlighting the importance of modeling multiple temporal scales within a trajectory.
Beyond architectural differences, a shared property of these recurrent models is that user behavior is summarized by a single evolving hidden state updated step-by-step, which can entangle heterogeneous behavioral signals when the same user exhibits distinct mobility patterns across contexts.
As a result, different behavioral processes, such as routine commuting, occasional exploration, and context-driven detours, are modeled under the same state transition dynamics, even though they may evolve at different temporal or spatial scales.

\textbf{Attention/Transformer and Graph-based Models.}
Attention mechanisms further improved POI recommendation by enabling flexible retrieval of relevant historical visits beyond adjacent steps.
STAN~\cite{luo2021stan} used spatiotemporal attention to capture correlations between non-adjacent check-ins and to mitigate overly strong recency bias.
GETNext~\cite{yang2022getnext} adopted self-attention to model long-range dependencies and leveraged global trajectory information to improve generalization.
Graph-based methods exploit relational structures among POIs and regions, e.g., STP-UDGAT~\cite{lim2020stpudgat}, Graph-Flashback~\cite{rao2022graphflashback}, and HMT-GRN~\cite{lim2022hmtgrn}, to propagate information over spatial or spatiotemporal graphs, capturing higher-order geographic correlations and alleviating sparsity via relational inductive bias.
Although these methods differ in how they access history (attention) or structure the space (graphs), they typically learn a unified user representation and apply a largely homogeneous evolution process for next-step prediction.
This design can limit adaptability under dynamic contexts where multiple coexisting mobility factors, such as stable routines, gradually shifting preferences, and rapidly changing exploratory intent, evolve with different spatiotemporal patterns.

\subsection{Spatiotemporal Modeling in Location-Based Services}

Human mobility exhibits strong regularities across multiple temporal scales (e.g., daily routines and weekly cycles) and is constrained by spatial feasibility and travel cost~\cite{yuan2013timeaware}.
To incorporate these signals, prior work typically injects time via discretized time-slot embeddings, periodic encodings, or learnable temporal embeddings, and models space through distance-aware transitions, coordinate features, or spatiotemporal graphs~\cite{liu2016strnn,luo2021stan}.
In practice, temporal modeling often needs to balance granularity and sparsity: fine-grained time bins preserve detailed periodic patterns but may suffer from data sparsity, while coarse bins improve robustness but may blur contextual distinctions.
Similarly, spatial modeling ranges from simple distance biases to richer geographic structures, trading off inductive bias and modeling flexibility.

While effective, many existing designs apply a largely uniform temporal or spatial modeling scheme to the entire user representation, implicitly assuming that different behavioral factors evolve at similar rates.
This assumption is misaligned with real-world mobility, where stable routines and rapidly changing intent coexist, and different behavioral components may respond differently to the same spatiotemporal context.
This mismatch motivates modeling heterogeneous spatiotemporal dynamics across decomposed behavioral components rather than enforcing a single transition regime.

\subsection{State Representation Learning}

\textbf{Disentanglement and Multi-Vector Representations.}
Disentangled representation learning~\cite{bengio2013representation,ma2019disentangled} aims to separate independent factors of variation and has been widely studied as a way to improve robustness and interpretability.
In sequential recommendation, multi-interest modeling extracts multiple vectors from user histories to represent diverse preferences and matches them to candidate items.
Such multi-vector representations are typically designed to capture preference diversity at the representation level, for example by clustering historical interactions or learning multiple attention heads over the same sequence.

\textbf{Limitations in Mobility Scenarios.}
These approaches are primarily developed for item recommendation and do not explicitly account for mobility-specific constraints such as spatial feasibility, temporal periodicity, and trajectory continuity.
More importantly, they focus on producing multiple preference vectors, but do not explicitly endow each vector with an independent state transition dynamic.
In POI recommendation, the key challenge is not only to represent diverse behavioral components, but also to track how each component evolves over time and space under different contexts.
Therefore, POI recommendation calls for multi-component state modeling where components can evolve under distinct spatiotemporal dynamics, rather than multi-vector representations that share a single evolution path.
This distinction is crucial: ADS-POI focuses on decomposing and tracking multiple latent decision states with independent dynamics, rather than merely extracting multiple preference embeddings for matching.

\subsection{Agentic Perspectives}

The agent abstraction~\cite{wooldridge1995agent} provides a conceptual lens for systems that maintain internal states and update them based on observations, and is often associated with reinforcement learning formulations~\cite{sutton2018rl}.
In recommender systems, the agent perspective is frequently used to emphasize sequential decision processes, but it is not inherently tied to policy optimization; it can also serve as a modeling viewpoint for structuring latent states and their interactions.
Our work adopts an agent-inspired perspective only as a modeling abstraction: ADS-POI is trained fully supervised and does not involve policy learning or environment interaction.
This perspective motivates decomposing user behavior into multiple latent sub-states that evolve independently and are selectively aggregated for prediction, resembling coordination among latent decision states rather than action selection via a learned policy.

\begin{figure*}
    \centering
    \includegraphics[width=1\linewidth]{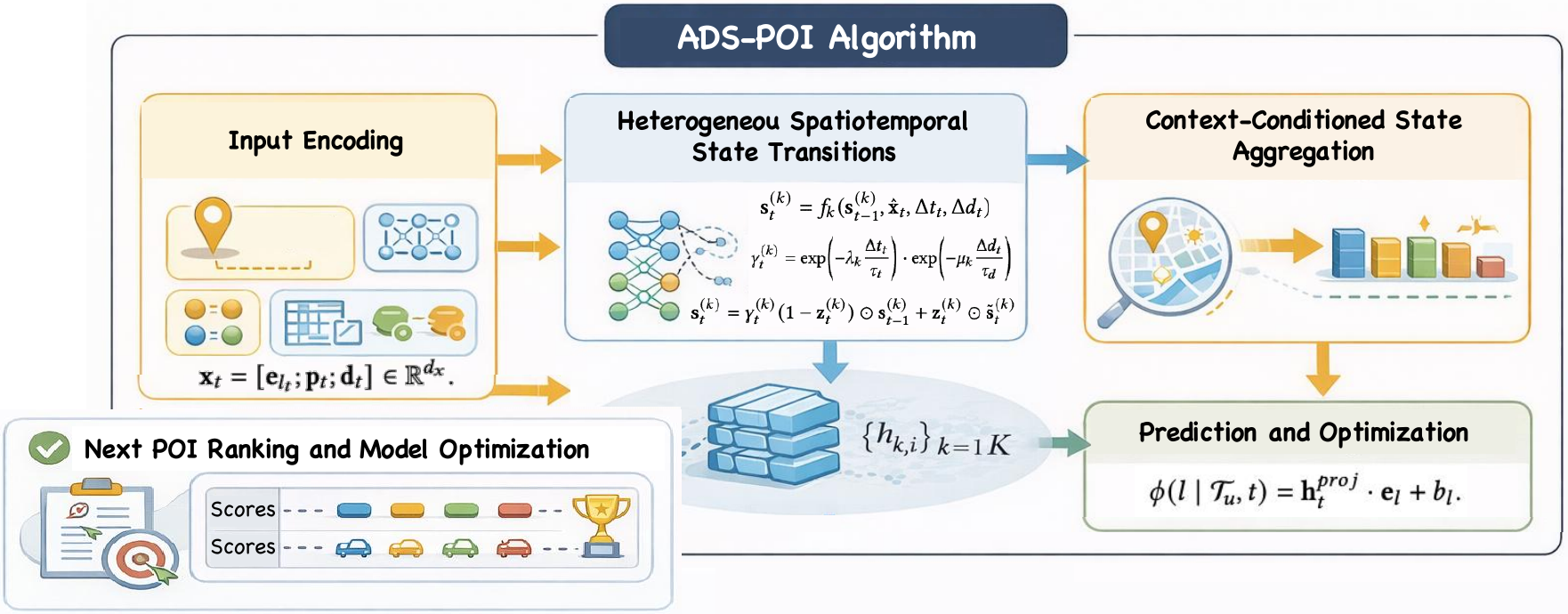}
    \caption{Algorithmic overview of ADS-POI. The model decomposes user behavior into multiple latent states with heterogeneous spatiotemporal transitions and performs context-adaptive state aggregation for next POI ranking.}
    \label{fig:overview}
\end{figure*}

\section{Problem Formulation}

We study the problem of next point-of-interest (POI) recommendation under a spatiotemporal sequential prediction setting.

\textbf{Point of Interest.}
A point of interest (POI) $l \in \mathcal{L}$ is a geographic location that users can visit, characterized by its coordinates $(lat_l, lon_l)$, semantic category, and optional attributes.
The set $\mathcal{L}$ denotes all POIs in the system.

\textbf{Check-in and User Trajectory.}
A check-in is defined as a tuple $(u, l, t)$, indicating that user $u$ visited POI $l$ at timestamp $t$.
For a given user $u$, a trajectory is a chronologically ordered sequence of check-ins:
\begin{equation}
\mathcal{T}_u = \{(l_1, t_1), (l_2, t_2), \ldots, (l_n, t_n)\},
\end{equation}
where $l_i \in \mathcal{L}$, $t_i \in \mathbb{T}$, and $t_1 < t_2 < \cdots < t_n$.
The trajectory $\mathcal{T}_u$ captures the historical mobility behavior of user $u$.

\textbf{Next POI Recommendation.}
Given a user trajectory $\mathcal{T}_u$ and a query time $t_{n+1}$, the goal of next POI recommendation is to predict the next location $l_{n+1}$ that the user will visit.
Formally, we aim to learn a scoring function
\begin{equation}
f: \mathcal{T}_u \times \mathcal{L} \times \mathbb{T} \rightarrow \mathbb{R},
\end{equation}
where $f(\mathcal{T}_u, l, t_{n+1})$ indicates the likelihood that user $u$ visits POI $l$ at time $t_{n+1}$.
At inference time, the recommendation is produced by ranking candidate POIs in $\mathcal{L}$ according to this score.

\textbf{Modeling Perspective.}
We adopt an agent-inspired modeling perspective to structure the user representation.
This perspective is used solely as a conceptual abstraction for organizing latent user states and their temporal evolution.
Our approach does not involve reinforcement learning, policy optimization, reward modeling, or environment interaction.
Instead, the model is trained in a purely supervised manner on historical check-in sequences, with the objective of maximizing next-step prediction accuracy.






\section{Method}

We propose ADS-POI, a spatiotemporal state decomposition framework for next POI recommendation.
The core idea is to represent a user with multiple parallel-evolving latent states, each capturing a distinct behavioral dimension with heterogeneous spatiotemporal dynamics.
Figure~\ref{fig:overview} illustrates the overall architecture.

\subsection{Multi-State User Representation}

At each step $i$, the user is represented by a set of $K$ latent sub-states:
\begin{equation}
\mathbf{s}_i = \{\mathbf{s}_i^{(1)}, \mathbf{s}_i^{(2)}, \ldots, \mathbf{s}_i^{(K)}\},
\end{equation}
where each sub-state $\mathbf{s}_i^{(k)} \in \mathbb{R}^{d_s}$ and the total state dimension is $d = K d_s$.
Instead of compressing all behavioral signals into a single latent vector, the multi-state formulation enables different behavioral aspects to be tracked and updated independently.
No explicit semantic meaning is imposed on individual sub-states.
Rather, their specialization emerges automatically during training, driven by heterogeneous state transition dynamics and context-conditioned aggregation.
This design alleviates state entanglement and provides greater modeling flexibility, while remaining fully data-driven.

\subsection{Input Encoding}

Each check-in is encoded by jointly modeling POI identity, temporal context, and spatial movement.

\paragraph{POI Encoding.}
Each POI $l$ is associated with a learnable embedding $\mathbf{e}_l \in \mathbb{R}^{d_e}$.

\paragraph{Temporal Encoding.}
We encode periodic temporal patterns using sinusoidal functions:
\begin{equation}
\mathbf{p}_i =
[\sin(\omega_h h_i), \cos(\omega_h h_i),
 \sin(\omega_w w_i), \cos(\omega_w w_i), \mathbf{e}_{slot}],
\end{equation}
where $h_i$ and $w_i$ denote the hour of day and the day of week at step $i$, and $\mathbf{e}_{slot}$ is a learnable time-slot embedding.
This design combines continuous periodic encoding with a discrete slot representation to capture both smooth temporal regularities and dataset-specific temporal patterns.

\paragraph{Spatial Encoding.}
Spatial movement is encoded using distance magnitude and direction:
\begin{equation}
\mathbf{d}_i = \mathbf{W}_d[\log(1+\Delta d_i), \mathbf{e}_{dist}, \Delta lat_i, \Delta lon_i],
\end{equation}
where $\Delta d_i$ is the geodesic distance between consecutive check-ins.
Here $\mathbf{e}_{dist}$ denotes a learnable embedding of distance buckets, and $\Delta lat_i, \Delta lon_i$ represent coordinate differences between successive locations.
The final input encoding at step $i$ is:
\begin{equation}
\mathbf{x}_i = [\mathbf{e}_{l_i}; \mathbf{p}_i; \mathbf{d}_i] \in \mathbb{R}^{d_x}.
\end{equation}

\paragraph{Dimension Alignment.}
When the input dimension $d_x$ differs from the sub-state dimension $d_s$, we apply a lightweight projection
$\Pi_x(\cdot)$ to map $\mathbf{x}_i$ into the sub-state space:
\begin{equation}
\hat{\mathbf{x}}_i = \Pi_x(\mathbf{x}_i), \quad \hat{\mathbf{x}}_i \in \mathbb{R}^{d_s}.
\end{equation}
Similarly, before scoring, the aggregated decision state is projected to match the POI embedding dimension:
\begin{equation}
\mathbf{h}_i^{proj} = \Pi_h(\mathbf{h}_i^{dec}), \quad \mathbf{h}_i^{proj} \in \mathbb{R}^{d_e}.
\end{equation}
These projections are used solely for dimension alignment and stable implementation, without altering the core model formulation.






\subsection{Heterogeneous Spatiotemporal State Transitions}

Different behavioral dimensions evolve at different temporal and spatial rates.
Accordingly, each sub-state is updated using an independent gated recurrent transition:
\begin{equation}
\mathbf{s}_i^{(k)} = f_k(\mathbf{s}_{i-1}^{(k)}, \hat{\mathbf{x}}_i, \Delta t_i, \Delta d_i),
\end{equation}
where $\Delta t_i$ and $\Delta d_i$ denote the time and distance gaps between consecutive check-ins.
We introduce state-specific spatiotemporal decay to modulate how each sub-state retains past information:
\begin{equation}
\gamma_i^{(k)} =
\exp\!\left(-\lambda_k \frac{\Delta t_i}{\tau_t}\right)
\cdot
\exp\!\left(-\mu_k \frac{\Delta d_i}{\tau_d}\right),
\end{equation}
where $\lambda_k$ and $\mu_k$ control the temporal and spatial sensitivity of sub-state $k$, and $\tau_t$ and $\tau_d$ are scaling constants used to normalize time and distance gaps.

To ensure valid decay behavior, $\lambda_k$ and $\mu_k$ are constrained to be non-negative.
In practice, we parameterize them with unconstrained scalars and apply a monotonic non-negative transform (e.g., \texttt{softplus}) during forward computation.

The final state update follows a GRU-style formulation:
\begin{equation}
\mathbf{s}_i^{(k)} =
\gamma_i^{(k)} (1-\mathbf{z}_i^{(k)}) \odot \mathbf{s}_{i-1}^{(k)}
+
\mathbf{z}_i^{(k)} \odot \tilde{\mathbf{s}}_i^{(k)},
\end{equation}
where $\mathbf{z}_i^{(k)}$ and $\tilde{\mathbf{s}}_i^{(k)}$ are the update gate and candidate state defined in a standard GRU.

\subsection{Context-Conditioned State Aggregation}

At prediction time, different internal states contribute unequally depending on the current spatiotemporal context.
Let $\mathbf{c}_i = [\mathbf{p}_i; \mathbf{d}_i]$ denote the context vector at step $i$.
The relevance of sub-state $k$ is computed as:
\begin{equation}
\alpha_i^{(k)} =
\frac{\exp(g(\mathbf{s}_i^{(k)}, \mathbf{c}_i)/\tau)}
{\sum_{j=1}^K \exp(g(\mathbf{s}_i^{(j)}, \mathbf{c}_i)/\tau)}.
\end{equation}

The aggregated decision state is obtained as a context-conditioned combination of sub-states:
\begin{equation}
\mathbf{h}_i^{dec} = \sum_{k=1}^K \alpha_i^{(k)} \mathbf{s}_i^{(k)}.
\end{equation}

\subsection{Prediction and Optimization}

\paragraph{Scoring.}
Given the aggregated decision state, the score of a candidate POI $l$ is computed as:
\begin{equation}
\phi(l \mid \mathcal{T}_u, i) = \mathbf{h}_i^{proj} \cdot \mathbf{e}_l + b_l,
\end{equation}
where $\mathbf{h}_i^{proj}$ denotes the projected decision state.

\paragraph{Sampled Softmax with Label Smoothing.}
For efficient training over large POI sets, we adopt sampled softmax with label smoothing:
\begin{equation}
\mathcal{L}_{\text{ce}}
=
-\sum_{j=0}^{N_{\text{neg}}}
y_j \log \big(\text{softmax}(\mathbf{z}_i)_j \big).
\end{equation}

\paragraph{Hard-negative BPR Margin Regularization.}
To further emphasize difficult negatives, we introduce a margin-based regularization over hard negatives:
\begin{equation}
\mathcal{L}_{\text{bpr}}
=
\frac{1}{K_h}\sum_{j\in \text{TopK}_h}
\max\big(0, m - (\phi(l^+) - \phi(l_j^-))\big).
\end{equation}

The overall training objective is:
\begin{equation}
\mathcal{L} = \mathcal{L}_{\text{ce}} + \beta \mathcal{L}_{\text{bpr}}.
\end{equation}

\subsection{Computational Complexity}

For a trajectory of length $n$, ADS-POI performs $K$ parallel sub-state transitions at each step, resulting in a time complexity of $O(n K d_s^2)$.
Sub-state transitions are independent and can be computed in parallel, making the model efficient in practice.
The detailed training procedure and optimization are provided in Algorithm~\ref{alg:ads-poi-simple} and the Appendix.



\begin{algorithm}[h]
\caption{ADS-POI: Overview}
\label{alg:ads-poi-simple}
\small
\begin{algorithmic}[1]
\REQUIRE User trajectory $\mathcal{T}_u=\{(l_1,t_1),\ldots,(l_n,t_n)\}$;
number of sub-states $K$.
\ENSURE Next-POI scores $\{\phi(l)\}_{l\in\mathcal{L}}$.

\STATE Initialize sub-states $\mathbf{s}_0^{(k)} \leftarrow \mathbf{0}$ for $k=1,\ldots,K$.
\FOR{$i=1$ to $n-1$}
  \STATE Encode check-in $\mathbf{x}_i \leftarrow \textsc{Encode}(l_i,t_i)$.
  \STATE Project input $\hat{\mathbf{x}}_i \leftarrow \Pi_x(\mathbf{x}_i)$.
  \FOR{$k=1$ to $K$}
    \STATE Compute decay $\gamma_i^{(k)} \leftarrow \textsc{Decay}_k(\Delta t_i,\Delta d_i)$.
    \STATE Update sub-state $\mathbf{s}_i^{(k)} \leftarrow \textsc{StateUpdate}_k(\mathbf{s}_{i-1}^{(k)},\hat{\mathbf{x}}_i,\gamma_i^{(k)})$.
  \ENDFOR
\ENDFOR
\STATE Compute context $\mathbf{c}_n \leftarrow \textsc{Context}(t_n,l_n)$.
\STATE Aggregate decision state $\mathbf{h}_n^{dec} \leftarrow \textsc{Aggregate}(\{\mathbf{s}_n^{(k)}\},\mathbf{c}_n)$.
\STATE Project decision state $\mathbf{h}_n^{proj} \leftarrow \Pi_h(\mathbf{h}_n^{dec})$.
\STATE Compute scores $\phi(l)\leftarrow \mathbf{h}_n^{proj}\cdot \mathbf{e}_l + b_l$ for all $l\in\mathcal{L}$.
\STATE \textbf{return} ranked POIs by $\phi(l)$.
\end{algorithmic}
\end{algorithm}

\section{Experiments}

\subsection{Experimental Setup}



Our experiments evaluate whether decomposing user behavior into multiple parallel spatiotemporal sub-states
improves next-POI ranking over conventional single-state modeling.
We focus on two practically challenging regimes: (i) heterogeneous mobility dynamics, where stable routines and rapidly changing preferences coexist, and (ii) large and sparse POI spaces, where limited observations increase ambiguity among plausible next locations.

\subsubsection{Datasets}

We evaluate ADS-POI on three widely-used real-world benchmark datasets: NYC and TKY from Foursquare~\cite{yang2015foursquare}, and CA from Gowalla~\cite{cho2011gowalla}.
We apply consistent preprocessing across datasets: (i) filter users with fewer than 10 check-ins, (ii) remove low-support POIs, (iii) remove consecutive duplicate check-ins at the same POI, and (iv) discard records with invalid timestamps or coordinates.
We adopt a leave-one-out split per user: the last check-in is used for testing, the second-to-last for validation, and the remaining check-ins for training.
All datasets exhibit high sparsity and long-tailed POI distributions after preprocessing, which is consistent with prior work on large-scale POI recommendation.
These datasets cover diverse mobility regimes.
NYC and TKY represent dense metropolitan environments with many geographically close POIs, where fine-grained spatiotemporal discrimination is important.
CA features a larger and sparser POI space with more dispersed mobility patterns, which allows us to assess robustness under different urban structures and movement ranges.

\begin{table*}[h]
\centering
\caption{Overall performance comparison under the full-ranking evaluation protocol. Best results are in \textbf{bold}, second best are \underline{underlined}. $\uparrow$ indicates higher is better.}
\label{tab:main}
\resizebox{\textwidth}{!}{
\begin{tabular}{l|ccccc|ccccc|ccccc}
\toprule
\multirow{2}{*}{\textbf{Method}} & \multicolumn{5}{c|}{\textbf{NYC (\%)}} & \multicolumn{5}{c|}{\textbf{TKY (\%)}} & \multicolumn{5}{c}{\textbf{CA (\%)}} \\ \cline{2-16}
~ & \textbf{HR@5$\uparrow$} & \textbf{HR@10$\uparrow$} & \textbf{NDCG@5$\uparrow$} & \textbf{NDCG@10$\uparrow$} & \textbf{MRR$\uparrow$} & \textbf{HR@5$\uparrow$} & \textbf{HR@10$\uparrow$} & \textbf{NDCG@5$\uparrow$} & \textbf{NDCG@10$\uparrow$} & \textbf{MRR$\uparrow$} & \textbf{HR@5$\uparrow$} & \textbf{HR@10$\uparrow$} & \textbf{NDCG@5$\uparrow$} & \textbf{NDCG@10$\uparrow$} & \textbf{MRR$\uparrow$} \\
\midrule
SASRec~\cite{kang2018sasrec} & 40.38 & 48.28 & 28.98 & 31.26 & 27.01 & 34.72 & 43.25 & 25.48 & 28.37 & 24.70 & 21.34 & 26.40 & 16.04 & 17.68 & 15.94 \\
BERT4Rec~\cite{sun2019bert4rec} & 39.60 & 46.70 & 29.03 & 30.94 & 27.11 & 36.17 & 43.47 & 26.70 & 29.08 & 25.39 & 23.22 & 28.22 & 17.26 & 18.66 & 16.61 \\
DuoRec~\cite{qiu2022duorec} & 34.72 & 40.69 & 24.54 & 26.34 & 22.76 & 24.66 & 31.19 & 18.77 & 20.89 & 18.67 & 19.47 & 24.04 & 14.47 & 15.85 & 14.31 \\
TiCoSeRec~\cite{dang2023uniform} & 40.69 & 48.08 & 29.65 & 32.04 & 27.54 & 29.33 & 37.04 & 21.83 & 24.21 & 21.51 & 22.66 & 28.11 & 16.99 & 18.48 & 16.75 \\
CrossDR~\cite{tao2023sttdr} & 40.80 & 47.20 & 30.67 & 32.59 & 28.66 & 32.51 & 40.57 & 23.58 & 26.19 & 22.65 & 22.11 & 28.61 & 16.11 & 18.18 & 15.79 \\
MAERec~\cite{ye2023maerec} & 37.65 & 44.53 & 27.72 & 29.85 & 26.10 & 32.73 & 40.66 & 23.93 & 26.14 & 23.05 & 24.37 & 29.37 & 17.76 & 19.48 & 17.09 \\
GETNext~\cite{yang2022getnext} & 43.60 & 51.30 & 32.56 & 35.13 & 30.97 & 41.11 & 48.50 & 30.75 & 33.21 & 29.21 & 24.26 & 29.32 & 17.83 & 19.48 & 17.16 \\
POIGDE~\cite{yang2024poigde} & 34.11 & 38.36 & 26.64 & 27.97 & 25.40 & 28.20 & 31.82 & 22.46 & 23.57 & 21.80 & 22.61 & 26.73 & 18.24 & 19.49 & 17.89 \\
MTNet~\cite{huang2024mtnet} & 47.57 & 53.24 & 35.85 & 37.69 & 33.40 & 44.15 & 51.22 & 32.99 & 35.37 & 31.18 & 30.25 & 36.08 & 22.39 & 24.27 & 21.35 \\
DiffuRec~\cite{li2024diffurec} & 37.25 & 40.79 & 29.16 & 30.23 & 27.74 & 34.50 & 39.89 & 26.89 & 28.70 & 25.94 & 21.67 & 24.15 & 17.35 & 18.06 & 16.60 \\
LLM4POI~\cite{li2024llm4poi} & 44.64 & 46.66 & 37.49 & 38.16 & 35.90 & 34.90 & 39.48 & 27.23 & 28.22 & 29.72 & 23.65 & 25.91 & 19.01 & 19.39 & 20.13 \\
CoMaPOI~\cite{zhong2025comapoi} & \underline{51.62} & \underline{59.01} & \underline{40.42} & \underline{42.82} & \underline{37.67} & \underline{45.83} & \underline{54.26} & \underline{34.48} & \underline{37.20} & \underline{31.82} & \underline{33.00} & \underline{39.16} & \underline{24.96} & \underline{26.96} & \underline{23.10} \\
\midrule
\rowcolor{lightblue}\textbf{ADS-POI} & \textbf{57.02} & \textbf{63.64} & \textbf{45.91} & \textbf{48.07} & \textbf{43.76} & \textbf{50.95} & \textbf{60.58} & \textbf{39.62} & \textbf{42.74} & \textbf{38.29} & \textbf{34.20} & \textbf{44.66} & \textbf{24.97} & \textbf{28.34} & \textbf{24.73} \\
\midrule
\rowcolor{lightblue}Improv. & +5.40\% & +4.63\% & +5.49\% & +5.25\% & +6.09\% & +5.12\% & +6.32\% & +5.14\% & +5.54\% & +6.47\% & +1.20\% & +5.50\% & +0.01\% & +1.38\% & +1.63\% \\
\bottomrule
\end{tabular}
}
\end{table*}

\subsubsection{Baselines}

We compare ADS-POI with a diverse set of strong baselines covering different modeling paradigms for next POI recommendation, as summarized in Table~\ref{tab:main}.
Sequential recommendation models, including SASRec~\cite{kang2018sasrec}, BERT4Rec~\cite{sun2019bert4rec}, DuoRec~\cite{qiu2022duorec}, TiCoSeRec~\cite{dang2023uniform}, CrossDR~\cite{tao2023sttdr}, and MAERec~\cite{ye2023maerec}, model user check-in sequences using recurrent or self-attention mechanisms, with various strategies to enhance robustness and representation quality.
Transformer-based and diffusion-based methods, such as GETNext~\cite{yang2022getnext} and DiffuRec~\cite{li2024diffurec}, capture long-range dependencies or generative uncertainty in sequential mobility patterns.
Graph-enhanced approaches, represented by POIGDE~\cite{yang2024poigde}, exploit relational structures among POIs to incorporate spatial correlations and alleviate data sparsity.
Trajectory-aware models, such as MTNet~\cite{huang2024mtnet}, explicitly model temporal preference evolution along mobility trajectories.
We also include recent LLM-based POI recommendation methods, including LLM4POI~\cite{li2024llm4poi} and CoMaPOI \cite{zhong2025comapoi}, which leverage large language models to encode user trajectories and semantic context.
We use official implementations whenever available; otherwise, we re-implement baselines following the original papers.
All hyperparameters are tuned on the validation set under a consistent search budget.
All methods are trained and evaluated on identical dataset splits and the same candidate set $\mathcal{L}$ to ensure fair comparison.



\subsubsection{Evaluation Protocol}

We use standard top-$k$ ranking metrics: Hit Rate (HR@$k$), Normalized Discounted Cumulative Gain (NDCG@$k$), and Mean Reciprocal Rank (MRR), with $k \in \{5,10\}$.
Unless otherwise specified, we adopt a full-ranking protocol: for each test instance, the model ranks all POIs in $\mathcal{L}$ and metrics are computed based on the rank of the ground-truth next POI.
We repeat each experiment with five random seeds and report mean and standard deviation.
We assess statistical significance using paired t-tests between ADS-POI and the strongest baseline on each dataset.



\subsubsection{Implementation Details}

ADS-POI uses embedding dimension $d=128$ and decomposes user representation into $K=4$ parallel sub-states (each with $d_s=32$).
We train ADS-POI with Adam (learning rate $10^{-3}$, batch size 256, dropout 0.2, weight decay $10^{-5}$) for up to 100 epochs with early stopping on validation MRR (patience 10).
We optimize a sampled softmax objective with label smoothing, using $N_{\text{neg}}=100$ negative POIs per step sampled from $\mathcal{L}$.
All experiments are conducted on GPUs under identical software environments across methods.

\begin{table*}[h]
\centering
\caption{Ablation study across three datasets. Notation: w/o = without; State Decomp. = State Decomposition; Agentic Aggr. = State Aggregation; Hetero. Trans. = Heterogeneous Transitions.}
\label{tab:ablation}
\resizebox{\textwidth}{!}{
\begin{tabular}{lccccc|ccccc|ccccc}
\toprule
\multirow{2}{*}{\textbf{Configuration}} 
& \multicolumn{5}{c|}{\textbf{NYC}} 
& \multicolumn{5}{c|}{\textbf{TKY}} 
& \multicolumn{5}{c}{\textbf{CA}} \\
\cline{2-16}
~ 
& \textbf{HR@5$\uparrow$} & \textbf{HR@10$\uparrow$} & \textbf{NDCG@5$\uparrow$} & \textbf{NDCG@10$\uparrow$} & \textbf{MRR$\uparrow$} & \textbf{HR@5$\uparrow$} & \textbf{HR@10$\uparrow$} & \textbf{NDCG@5$\uparrow$} & \textbf{NDCG@10$\uparrow$} & \textbf{MRR$\uparrow$} & \textbf{HR@5$\uparrow$} & \textbf{HR@10$\uparrow$} & \textbf{NDCG@5$\uparrow$} & \textbf{NDCG@10$\uparrow$} & \textbf{MRR$\uparrow$} \\
\midrule
\rowcolor{lightblue}
\textbf{ADS-POI (Full)} 
& \textbf{57.02} & \underline{63.64} & \textbf{45.91} & \textbf{48.07} & \textbf{43.76} & \textbf{50.95} & \textbf{60.58} & \textbf{39.62} & \textbf{42.74} & \textbf{38.29}
& \textbf{34.20} & \textbf{44.66} & \textbf{24.97} & \textbf{28.34} & \textbf{24.73} \\
\midrule
w/o State Decomp. 
& 56.32 & \textbf{64.62} & 42.10 & 44.81 & 39.12
& 29.21 & 37.31 & 20.46 & 23.09 & 19.80
& 30.11 & 39.85 & 21.12 & 24.27 & 20.33 \\
Single State ($K{=}1$) 
& 50.78 & 59.26 & 37.74 & 40.53 & 35.34
& 27.30 & 35.37 & 19.33 & 21.94 & 18.95
& 29.28 & 38.61 & 20.34 & 23.32 & 19.62 \\
w/o Agentic Aggr. 
& 52.74 & 59.81 & 39.53 & 41.84 & 36.82
& 28.57 & 36.78 & 20.48 & 23.13 & 20.07
& 29.85 & 39.22 & 20.81 & 23.89 & 20.18 \\
w/o Hetero. Trans. 
& \underline{56.67} & 63.44 & \underline{45.59} & \underline{47.80} & \underline{43.50}
& \underline{47.69} & \underline{57.49} & \underline{37.03} & \underline{40.21} & \underline{35.95}
& \underline{33.65} & \underline{43.84} & \underline{24.29} & \underline{27.69} & \underline{24.11} \\
\bottomrule
\end{tabular}
}
\end{table*}

\begin{table}[h]
\centering
\caption{Efficiency comparison. Latency/throughput are measured under the same input setting with batch size 1. FLOPs are estimated under the same setting.}
\label{tab:efficiency}
\resizebox{0.7\linewidth}{!}{
\begin{tabular}{lcccc}
\toprule
\multirow{2}{*}{\textbf{Method}} 
& \textbf{Latency} 
& \textbf{Throughput} 
& \textbf{Memory} 
& \textbf{FLOPs} \\
~ 
& \textbf{(ms)$\downarrow$} 
& \textbf{(queries/s)$\uparrow$} 
& \textbf{(MB)$\downarrow$} 
& \textbf{(GFLOPs)$\downarrow$} \\
\midrule
SASRec   & 2.5  & 420 & 128 & 0.3 \\
GETNext  & 8.5           & 125          & 312          & 1.1 \\
MTNet    & 6.8 & 155 & 256 & 0.9 \\
CoMaPOI  & 156.0         & 6.5          & 8,500        & 150.0 \\
\rowcolor{lightblue}\textbf{ADS-POI} 
         & 22.0 & 45 & 768 & 6.5 \\
\bottomrule
\end{tabular}
}
\end{table}

\begin{figure*}[h]
\centering
\includegraphics[width=0.75\linewidth]{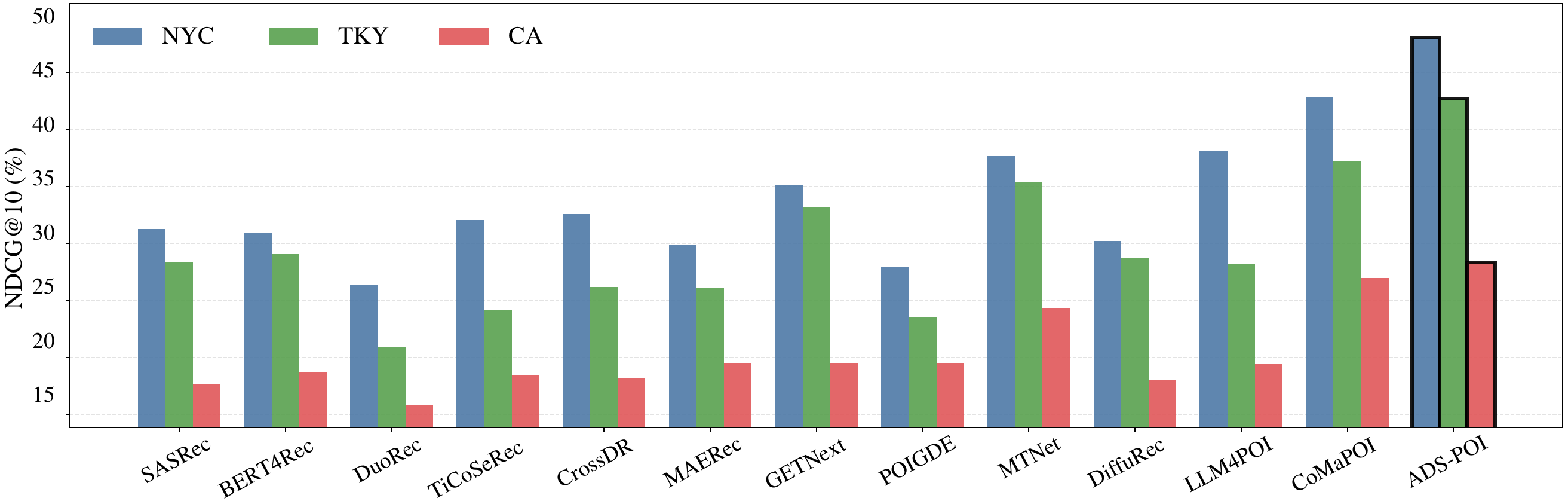}
\caption{Overall performance comparison on three datasets under full-ranking evaluation. ADS-POI consistently outperforms strong baselines in HR@10 and NDCG@10.}
\label{fig:overall}
\end{figure*}

\begin{figure*}[h]
\centering
\includegraphics[width=0.75\linewidth]{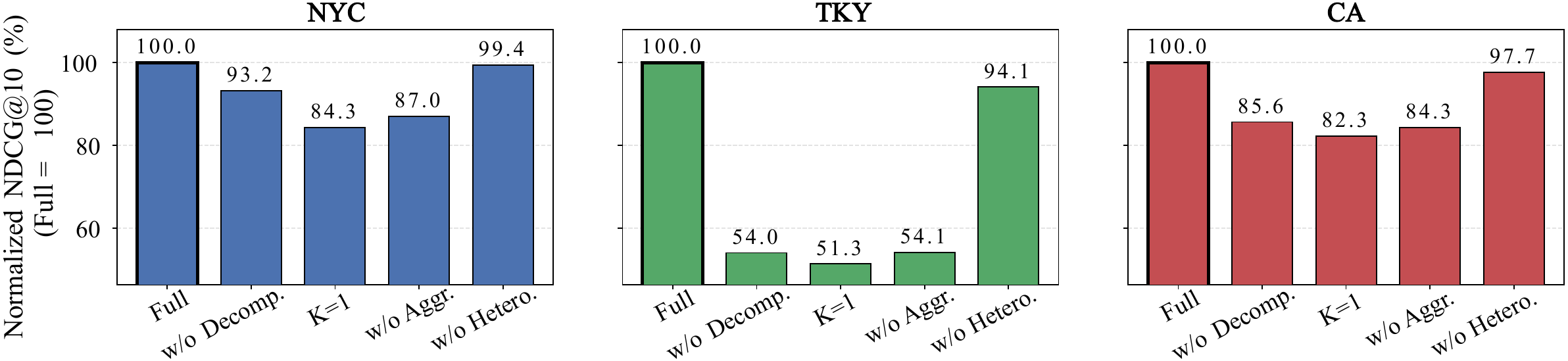}
\caption{Normalized impact of different components in ADS-POI on NYC. Performance is normalized by the full model (100\%). State decomposition and context-conditioned aggregation contribute most to the gains, while heterogeneous transitions provide additional improvements.}
\label{fig:ablation_impact}
\end{figure*}

\begin{figure}[h]
\centering
\includegraphics[width=0.7\columnwidth]{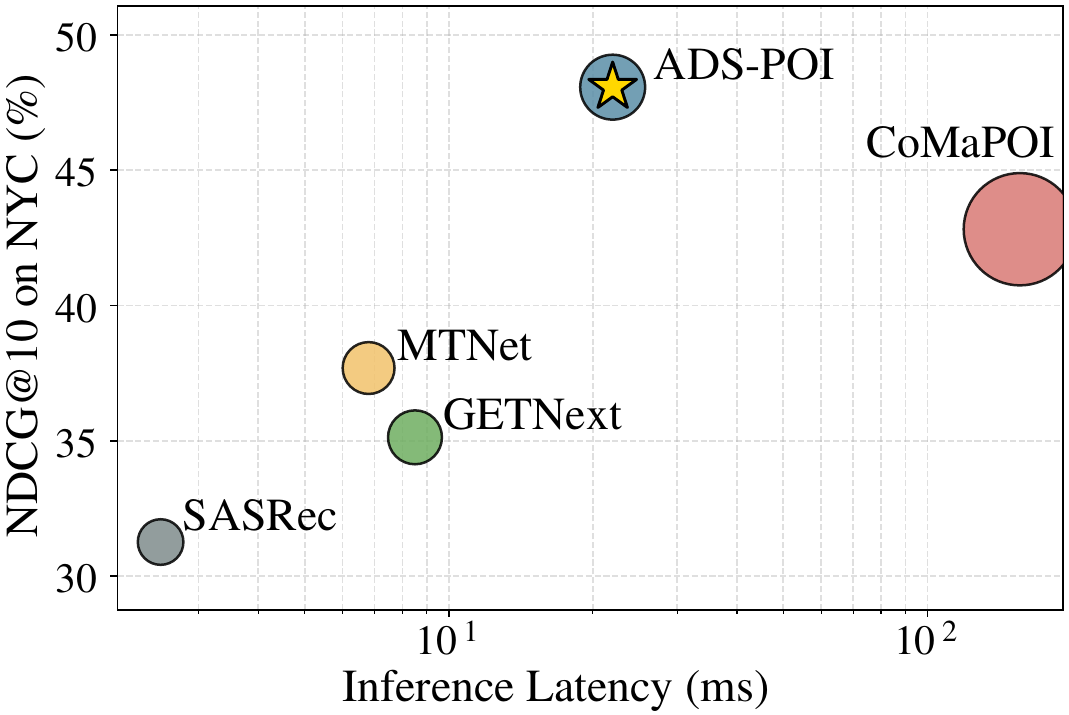}
\caption{Accuracy--efficiency trade-off. ADS-POI achieves a favorable balance between ranking performance and computational cost.}
\label{fig:tradeoff}
\end{figure}

\subsection{Overall Performance}

Table~\ref{tab:main} reports the performance comparison under the full-ranking evaluation protocol.
Across all three datasets and evaluation metrics, ADS-POI consistently outperforms strong baselines,
demonstrating the effectiveness of spatiotemporal state decomposition for next POI recommendation.
The improvements are observed on both hit-based metrics (HR@$k$) and ranking-sensitive metrics (NDCG@$k$ and MRR),
indicating that ADS-POI not only increases the likelihood of retrieving the correct next POI,
but also improves its ranking at top positions, which is critical for practical recommendation scenarios.
Notably, ADS-POI achieves clear gains over competitive sequential, graph-based, and LLM-based methods,
including MTNet and CoMaPOI.

Performance gains are consistent across datasets with diverse mobility characteristics.
On dense metropolitan datasets such as NYC and TKY, where many POIs are geographically close and highly competitive,
ADS-POI shows particularly strong improvements, suggesting that modeling multiple parallel states helps resolve fine-grained spatiotemporal ambiguity.
On the CA dataset, which exhibits more dispersed POIs and longer-range movements,
ADS-POI also maintains a clear advantage, demonstrating robustness across different spatial scales and sparsity regimes.
Compared with recent LLM-based approaches, which typically rely on a trajectory-level representation without explicitly decomposing heterogeneous spatiotemporal dynamics,
ADS-POI models multiple parallel sub-states with distinct decay behaviors and aggregates them in a context-conditioned manner.
This design enables more precise alignment between historical mobility patterns and the current decision context,
leading to consistent improvements across datasets.
Figure~\ref{fig:overall} further visualizes these trends using HR@10 and NDCG@10.

\subsection{Ablation Study}

We examine the contribution of key components in ADS-POI through an ablation study that selectively removes individual design elements.
Table~\ref{tab:ablation} reports results across three datasets.
Removing state decomposition (w/o State Decomp.) leads to noticeable degradation in ranking performance, especially on NDCG and MRR, indicating that explicitly separating heterogeneous behavioral factors is important for next-POI recommendation.
The single-state variant ($K{=}1$) performs even worse, confirming that maintaining multiple parallel sub-states provides additional modeling capacity beyond a single recurrent representation.

Disabling context-conditioned aggregation (w/o Agentic Aggr.) also results in clear performance drops, showing that dynamically coordinating sub-states according to the current spatiotemporal context is crucial for effective ranking.
In contrast, removing heterogeneous transitions (w/o Hetero. Trans.) causes smaller but consistent degradation, suggesting that state-specific spatiotemporal decay mainly contributes by improving robustness and specialization rather than dominating performance gains.
Figure~\ref{fig:ablation_impact} illustrates the relative impact of each component on NYC.
Although certain ablations may slightly improve an individual metric on a specific dataset,
the full ADS-POI model achieves the strongest ranking quality across metrics,
particularly on NDCG and MRR, which place greater emphasis on top-ranked positions.

\subsection{Efficiency Analysis}

Table~\ref{tab:efficiency} compares computational efficiency in terms of latency, throughput, memory consumption, and estimated FLOPs under the same input setting.
Maintaining $K$ parallel sub-states introduces additional computation compared with lightweight transformer baselines.
Nevertheless, ADS-POI remains substantially more efficient than CoMaPOI, which incurs much higher latency, memory footprint, and computational cost.

ADS-POI avoids expensive full self-attention over long sequences and large-scale graph construction or message passing.
Instead, it relies on parallel recurrent-style updates that can be efficiently implemented, enabling strong ranking performance without resorting to extremely expensive model scaling.
This design results in a favorable balance between effectiveness and efficiency.
In contrast, the high computational overhead of CoMaPOI limits its applicability in latency-sensitive or resource-constrained scenarios.
Despite its lower computational cost relative to such LLM-based approaches, ADS-POI consistently achieves better ranking accuracy, highlighting the benefit of targeted architectural inductive biases.
Figure~\ref{fig:tradeoff} illustrates this accuracy--efficiency trade-off.







\subsection{Robustness to Hyperparameters}

We analyze the robustness of ADS-POI with respect to key hyperparameters, including the number of sub-states $K$ and the embedding dimension $d$.
Rather than tuning for peak performance, this analysis evaluates whether the proposed framework is sensitive to specific parameter choices.
Introducing multiple sub-states consistently improves performance over the single-state variant, confirming the importance of state decomposition for modeling heterogeneous mobility patterns.
As $K$ increases, performance quickly stabilizes once a moderate number of sub-states is used, indicating that ADS-POI does not rely on a finely tuned choice of $K$.
A similar trend is observed for the embedding dimension: increasing $d$ improves performance at small scales, while gains gradually saturate beyond a moderate dimension, reflecting diminishing returns from additional capacity.
These results show that ADS-POI is robust across a wide and practical range of hyperparameter settings, and strong performance can be achieved without extensive tuning.
Detailed results are provided in Appendix~\ref{sec_appendix_k}.





\subsection{Discussion and Practical Implications}

The robustness of ADS-POI has direct implications for real-world deployment.
In large-scale location-based services, extensive hyperparameter tuning is often impractical due to limited computational budgets and continuously evolving data distributions.
Our results show that ADS-POI achieves strong performance across a broad and reasonable range of hyperparameter settings, making it well suited to such production environments.

From a practical perspective, the limited sensitivity to the number of sub-states means that practitioners do not need to carefully balance model complexity against performance when adopting state decomposition.
Likewise, the saturation behavior with respect to embedding dimension suggests that moderate-sized embeddings are sufficient, which helps control memory consumption and inference latency.
Together, these properties make ADS-POI a reliable and deployable solution for next POI recommendation in dynamic settings, where robustness, scalability, and ease of use are often as important as peak predictive accuracy.

\section{Conclusion}

This paper proposed ADS-POI, a spatiotemporal state decomposition framework for next POI recommendation.
By representing user behavior with multiple parallel latent states that evolve under heterogeneous spatiotemporal dynamics, ADS-POI alleviates the entanglement inherent in single-state modeling and enables more flexible mobility representation.
A context-conditioned aggregation mechanism further allows the model to adaptively emphasize relevant behavioral components for each prediction.
Extensive experiments on three real-world benchmark datasets demonstrate consistent improvements over strong baselines under a full-ranking evaluation protocol, across both dense and sparse POI environments.

\clearpage
\bibliographystyle{ACM-Reference-Format}
\bibliography{main2}

@inproceedings{rendle2010fpmc,
  title={Factorizing Personalized {Markov} Chains for Next-basket Recommendation},
  author={Rendle, Steffen and Freudenthaler, Christoph and Schmidt-Thieme, Lars},
  booktitle={Proceedings of the 19th International Conference on World Wide Web},
  pages={811--820},
  year={2010},
  organization={ACM}
}

@inproceedings{feng2015prme,
  title={Personalized Ranking Metric Embedding for Next New {POI} Recommendation},
  author={Feng, Shanshan and Li, Xutao and Zeng, Yifeng and Cong, Gao and Chee, Yeow Meng and Yuan, Quan},
  booktitle={Proceedings of the 24th International Joint Conference on Artificial Intelligence},
  pages={2069--2075},
  year={2015},
  organization={AAAI Press}
}

@inproceedings{liu2016strnn,
  title={Predicting the Next Location: A Recurrent Model with Spatial and Temporal Contexts},
  author={Liu, Qiang and Wu, Shu and Wang, Liang and Tan, Tieniu},
  booktitle={Proceedings of the AAAI Conference on Artificial Intelligence},
  volume={30},
  number={1},
  pages={194--200},
  year={2016}
}

@inproceedings{sun2020lstpm,
  title={Where to Go Next: Modeling Long- and Short-Term User Preferences for Point-of-Interest Recommendation},
  author={Sun, Ke and Qian, Tieyun and Chen, Tong and Liang, Yile and Nguyen, Quoc Viet Hung and Yin, Hongzhi},
  booktitle={Proceedings of the AAAI Conference on Artificial Intelligence},
  volume={34},
  number={1},
  pages={214--221},
  year={2020}
}

@inproceedings{luo2021stan,
  title={{STAN}: Spatio-Temporal Attention Network for Next Location Recommendation},
  author={Luo, Yingtao and Liu, Qiang and Liu, Zhaocheng},
  booktitle={Proceedings of the Web Conference 2021},
  pages={2177--2185},
  year={2021},
  organization={ACM}
}

@inproceedings{lim2020stpudgat,
  title={{STP-UDGAT}: Spatial-Temporal-Preference User Dimensional Graph Attention Network for Next {POI} Recommendation},
  author={Lim, Nicholas and Hooi, Bryan and Ng, See-Kiong and Wang, Xueou and Goh, Yong Liang and Weng, Renrong and Varadarajan, Jagannadan},
  booktitle={Proceedings of the 29th ACM International Conference on Information and Knowledge Management},
  pages={845--854},
  year={2020},
  organization={ACM}
}

@inproceedings{rao2022graphflashback,
  title={Graph-Flashback Network for Next Location Recommendation},
  author={Rao, Xuan and Chen, Lisi and Liu, Yong and Shang, Shuo and Yao, Bin and Han, Peng},
  booktitle={Proceedings of the 28th ACM SIGKDD Conference on Knowledge Discovery and Data Mining},
  pages={1463--1471},
  year={2022},
  organization={ACM}
}

@inproceedings{lim2022hmtgrn,
  title={Hierarchical Multi-Task Graph Recurrent Network for Next {POI} Recommendation},
  author={Lim, Nicholas and Hooi, Bryan and Ng, See-Kiong and Goh, Yong Liang and Weng, Renrong and Tan, Rui},
  booktitle={Proceedings of the 45th International ACM SIGIR Conference on Research and Development in Information Retrieval},
  pages={1133--1143},
  year={2022},
  organization={ACM}
}

@article{yang2015foursquare,
  title={Modeling User Activity Preference by Leveraging User Spatial Temporal Characteristics in {LBSNs}},
  author={Yang, Dingqi and Zhang, Daqing and Zheng, Vincent W. and Yu, Zhiyong},
  journal={IEEE Transactions on Systems, Man, and Cybernetics: Systems},
  volume={45},
  number={1},
  pages={129--142},
  year={2015},
  publisher={IEEE}
}

@inproceedings{cho2011gowalla,
  title={Friendship and Mobility: User Movement in Location-Based Social Networks},
  author={Cho, Eunjoon and Myers, Seth A. and Leskovec, Jure},
  booktitle={Proceedings of the 17th ACM SIGKDD International Conference on Knowledge Discovery and Data Mining},
  pages={1082--1090},
  year={2011},
  organization={ACM}
}

@article{yuan2013timeaware,
  title={Time-aware Point-of-interest Recommendation},
  author={Yuan, Quan and Cong, Gao and Ma, Zongyang and Sun, Aixin and Thalmann, Nadia Magnenat},
  journal={Proceedings of the 36th International ACM SIGIR Conference on Research and Development in Information Retrieval},
  pages={363--372},
  year={2013}
}

@inproceedings{ma2019disentangled,
  title={Learning Disentangled Representations for Recommendation},
  author={Ma, Jianxin and Zhou, Chang and Cui, Peng and Yang, Hongxia and Zhu, Wenwu},
  booktitle={Advances in Neural Information Processing Systems},
  volume={32},
  year={2019}
}

@article{bengio2013representation,
  title={Representation Learning: A Review and New Perspectives},
  author={Bengio, Yoshua and Courville, Aaron and Vincent, Pascal},
  journal={IEEE Transactions on Pattern Analysis and Machine Intelligence},
  volume={35},
  number={8},
  pages={1798--1828},
  year={2013},
  publisher={IEEE}
}

@article{wooldridge1995agent,
  title={Intelligent Agents: Theory and Practice},
  author={Wooldridge, Michael and Jennings, Nicholas R.},
  journal={The Knowledge Engineering Review},
  volume={10},
  number={2},
  pages={115--152},
  year={1995},
  publisher={Cambridge University Press}
}

@book{sutton2018rl,
  title={Reinforcement Learning: An Introduction},
  author={Sutton, Richard S. and Barto, Andrew G.},
  year={2018},
  publisher={MIT Press},
  edition={2nd}
}

@inproceedings{kang2018sasrec,
  title={Self-attentive sequential recommendation},
  author={Kang, Wang-Cheng and McAuley, Julian},
  booktitle={2018 IEEE international conference on data mining (ICDM)},
  pages={197--206},
  year={2018},
  organization={IEEE}
}

@inproceedings{sun2019bert4rec,
  title={BERT4Rec: Sequential recommendation with bidirectional encoder representations from transformer},
  author={Sun, Fei and Liu, Jun and Wu, Jian and Pei, Changhua and Lin, Xiao and Ou, Wenwu and Jiang, Peng},
  booktitle={Proceedings of the 28th ACM international conference on information and knowledge management},
  pages={1441--1450},
  year={2019}
}

@inproceedings{dang2023uniform,
  title={Uniform sequence better: Time interval aware data augmentation for sequential recommendation},
  author={Dang, Yizhou and Yang, Enneng and Guo, Guibing and Jiang, Linying and Wang, Xingwei and Xu, Xiaoxiao and Sun, Qinghui and Liu, Hong},
  booktitle={Proceedings of the AAAI conference on artificial intelligence},
  volume={37},
  number={4},
  pages={4225--4232},
  year={2023}
}

@inproceedings{yang2022getnext,
  title={GETNext: Trajectory flow map enhanced transformer for next POI recommendation},
  author={Yang, Song and Liu, Jiamou and Zhao, Kaiqi},
  booktitle={Proceedings of the 45th International ACM SIGIR Conference on research and development in information retrieval},
  pages={1144--1153},
  year={2022}
}

@inproceedings{tao2023sttdr,
  title={Next POI recommendation based on spatial and temporal disentanglement representation},
  author={Tao, Hongjin and Zeng, Jun and Wang, Ziwei and Gao, Min and Wen, Junhao},
  booktitle={2023 IEEE International Conference on Web Services (ICWS)},
  pages={84--90},
  year={2023},
  organization={IEEE}
}

@inproceedings{ye2023maerec,
  title={Graph masked autoencoder for sequential recommendation},
  author={Ye, Yaowen and Xia, Lianghao and Huang, Chao},
  booktitle={Proceedings of the 46th international ACM SIGIR conference on research and development in information retrieval},
  pages={321--330},
  year={2023}
}

@inproceedings{qiu2022duorec,
  title={Contrastive learning for representation degeneration problem in sequential recommendation},
  author={Qiu, Ruihong and Huang, Zi and Yin, Hongzhi and Wang, Zijian},
  booktitle={Proceedings of the fifteenth ACM international conference on web search and data mining},
  pages={813--823},
  year={2022}
}

@article{yang2024poigde,
  title={Siamese learning based on graph differential equation for Next-POI recommendation},
  author={Yang, Yuxuan and Zhou, Siyuan and Weng, He and Wang, Dongjing and Zhang, Xin and Yu, Dongjin and Deng, Shuiguang},
  journal={Applied Soft Computing},
  volume={150},
  pages={111086},
  year={2024},
  publisher={Elsevier}
}

@inproceedings{huang2024mtnet,
  title={Learning time slot preferences via mobility tree for next poi recommendation},
  author={Huang, Tianhao and Pan, Xuan and Cai, Xiangrui and Zhang, Ying and Yuan, Xiaojie},
  booktitle={Proceedings of the AAAI Conference on Artificial Intelligence},
  volume={38},
  number={8},
  pages={8535--8543},
  year={2024}
}

@article{li2024diffurec,
  title={Diffurec: A diffusion model for sequential recommendation},
  author={Li, Zihao and Sun, Aixin and Li, Chenliang},
  journal={ACM Transactions on Information Systems},
  volume={42},
  number={3},
  pages={1--28},
  year={2023},
  publisher={ACM New York, NY}
}

@inproceedings{li2024llm4poi,
  title={Large language models for next point-of-interest recommendation},
  author={Li, Peibo and de Rijke, Maarten and Xue, Hao and Ao, Shuang and Song, Yang and Salim, Flora D},
  booktitle={Proceedings of the 47th International ACM SIGIR Conference on Research and Development in Information Retrieval},
  pages={1463--1472},
  year={2024}
}

@inproceedings{zhong2025comapoi,
  title={Comapoi: A collaborative multi-agent framework for next poi prediction bridging the gap between trajectory and language},
  author={Zhong, Lin and Wang, Lingzhi and Yang, Xu and Liao, Qing},
  booktitle={Proceedings of the 48th International ACM SIGIR Conference on Research and Development in Information Retrieval},
  pages={1768--1778},
  year={2025}
}

\clearpage
\appendix
\onecolumn

\section{Algorithm}

\begin{algorithm}[h]
\caption{ADS-POI: Training and Inference}
\label{alg:ads-poi}
\small
\begin{algorithmic}[1]
\REQUIRE POI set $\mathcal{L}$; user trajectories $\{\mathcal{T}_u\}$;
number of sub-states $K$; sub-state dimension $d_s$; POI embedding dimension $d_e$;
temperature $\tau$; decay normalizers $(\tau_t,\tau_d)$;
number of negatives $N_{\text{neg}}$; label smoothing $\epsilon$;
number of hard negatives $K_h$; margin $m$; BPR weight $\beta$;
negative sampler $Q(\cdot)$.
\ENSURE Model parameters $\Theta$.

\vspace{2pt}
\STATE Initialize POI embeddings $\{\mathbf{e}_l\}_{l\in\mathcal{L}}$, biases $\{b_l\}$, and all network parameters $\Theta$.

\FOR{each training epoch}
  \FOR{each mini-batch of users}
    \FOR{each user $u$ in the mini-batch}
      \STATE Initialize sub-states $\mathbf{s}_0^{(k)} \leftarrow \mathbf{0}$ for $k=1,\dots,K$.
      \FOR{$i=1$ to $n_u-1$} 
        \STATE Compute time gap $\Delta t_i \leftarrow t_i - t_{i-1}$ and distance gap $\Delta d_i \leftarrow \textsc{GeoDist}(l_i, l_{i-1})$.
        \STATE Encode input $\mathbf{x}_i \leftarrow [\mathbf{e}_{l_i}; \mathbf{p}_i; \mathbf{d}_i]$.
        \STATE Project input $\hat{\mathbf{x}}_i \leftarrow \Pi_x(\mathbf{x}_i)$.
        \FOR{$k=1$ to $K$}
          \STATE Compute decay $\gamma_i^{(k)} \leftarrow \exp(-\lambda_k \Delta t_i / \tau_t)\cdot \exp(-\mu_k \Delta d_i / \tau_d)$.
          \STATE Update sub-state $\mathbf{s}_i^{(k)} \leftarrow \textsc{StateUpdate}_k(\mathbf{s}_{i-1}^{(k)}, \hat{\mathbf{x}}_i, \gamma_i^{(k)})$.
        \ENDFOR

        \STATE Form context $\mathbf{c}_i \leftarrow [\mathbf{p}_i; \mathbf{d}_i]$.
        \STATE Compute aggregation weights $\alpha_i^{(k)} \leftarrow \textsc{Softmax}(g(\mathbf{s}_i^{(k)}, \mathbf{c}_i)/\tau)$.
        \STATE Aggregate decision state $\mathbf{h}_i^{dec} \leftarrow \sum_{k=1}^K \alpha_i^{(k)} \mathbf{s}_i^{(k)}$.
        \STATE Project decision state $\mathbf{h}_i^{proj} \leftarrow \Pi_h(\mathbf{h}_i^{dec})$.

        \STATE Positive next POI $l^+ \leftarrow l_{i+1}$.
        \STATE Sample negatives $\{l_j^-\}_{j=1}^{N_{\text{neg}}} \sim Q(\cdot)$ and form candidate set $\mathcal{C}_i$.

        \STATE Compute logits $\mathbf{z}_i[j] \leftarrow \mathbf{h}_i^{proj}\cdot \mathbf{e}_{l_j} + b_{l_j}$ for all $l_j \in \mathcal{C}_i$.

        \STATE Construct smoothed label $\mathbf{y}$ with $y_0 = 1-\epsilon + \epsilon/|\mathcal{C}_i|$ and $y_j=\epsilon/|\mathcal{C}_i|$ for $j\neq 0$.
        \STATE $\mathcal{L}_{\text{ce}} \leftarrow -\sum_{j} y_j \log(\textsc{Softmax}(\mathbf{z}_i)[j])$.

        \STATE Select hardest negatives $\mathcal{H}\leftarrow \textsc{TopK}_h(\{\mathbf{z}_i[j]\}_{j>0})$.
        \STATE $\mathcal{L}_{\text{bpr}} \leftarrow \frac{1}{K_h}\sum_{j\in\mathcal{H}} \max\big(0,\, m - (\mathbf{z}_i[0] - \mathbf{z}_i[j])\big)$.
        \STATE Accumulate loss $\mathcal{L} \leftarrow \mathcal{L}_{\text{ce}} + \beta \mathcal{L}_{\text{bpr}}$.
      \ENDFOR
    \ENDFOR
    \STATE Update $\Theta$ by Adam on the batch loss.
  \ENDFOR
\ENDFOR

\vspace{2pt}
\STATE \textbf{Inference:} Given a trajectory prefix up to step $i$, compute $\mathbf{h}_i^{proj}$ and rank all POIs
$\phi(l)=\mathbf{h}_i^{proj}\cdot \mathbf{e}_l + b_l$, then recommend top-$K$ POIs.
\end{algorithmic}
\end{algorithm}

\section{Parameter Sensitivity Analysis}
\label{sec_appendix_k}

We analyze the sensitivity of ADS-POI to key hyperparameters,
including the number of sub-states $K$ and the embedding dimension $d$.
These experiments aim to assess robustness rather than to exhaustively tune hyperparameters.

\paragraph{Effect of the Number of Sub-states $K$.}
Table~\ref{tab:param_sensitivity_K} reports results when varying the number of sub-states $K$.
In this experiment, we keep the \emph{sub-state dimension} $d_s$ fixed and thus the total embedding dimension scales with $K$ (i.e., $d = K \cdot d_s$).
When $K=1$, ADS-POI degenerates to a single-state variant and yields noticeably worse ranking quality,
suggesting that a unified representation is insufficient to capture heterogeneous mobility patterns.
Increasing $K$ consistently improves performance and the results become relatively stable for moderate to large $K$ values.
Overall, ADS-POI is not overly sensitive to the exact choice of $K$ once $K$ is moderately large.

\paragraph{Effect of Embedding Dimension $d$.}
Table~\ref{tab:param_sensitivity_d} evaluates the impact of the embedding dimension $d$ under the single-state setting ($K=1$),
so that the effect of representational capacity can be assessed in isolation.
We observe that performance improves when increasing $d$ from smaller values, and then gradually plateaus at larger dimensions,
indicating diminishing returns from further enlarging embeddings.

Overall, these results suggest that ADS-POI is robust to hyperparameter choices in a wide range,
and does not require careful tuning to obtain strong performance.

\begin{table}[h]
\centering
\caption{Parameter sensitivity of ADS-POI w.r.t. the number of sub-states $K$ on three datasets. Best results are in \textbf{bold}, and second-best results are \underline{underlined}.}
\label{tab:param_sensitivity_K}
\small
\begin{tabular}{c|ccc|ccc|ccc}
\toprule
\multirow{2}{*}{$K$} 
& \multicolumn{3}{c|}{\textbf{NYC}} 
& \multicolumn{3}{c|}{\textbf{TKY}} 
& \multicolumn{3}{c}{\textbf{CA}} \\
\cmidrule(lr){2-4} \cmidrule(lr){5-7} \cmidrule(lr){8-10}
& \textbf{HR@10} & \textbf{NDCG@10} & \textbf{MRR}
& \textbf{HR@10} & \textbf{NDCG@10} & \textbf{MRR}
& \textbf{HR@10} & \textbf{NDCG@10} & \textbf{MRR} \\
\midrule
1 & \underline{64.62} & 44.81 & 39.12
  & 58.43 & 40.27 & 35.18
  & 55.87 & 38.64 & 33.47 \\
2 & 61.10 & 45.58 & 41.28
  & 56.92 & 41.06 & 36.61
  & 54.18 & 39.37 & 34.72 \\
3 & 63.24 & 48.07 & 43.91
  & 59.08 & 43.84 & 38.93
  & 56.29 & 41.76 & 36.92 \\
4 & 63.64 & 48.07 & 43.76
  & \underline{59.21} & 43.79 & 38.88
  & 56.41 & 41.69 & 36.84 \\
5 & 63.58 & 48.46 & 44.26
  & 59.17 & 44.16 & 39.24
  & 56.36 & 42.09 & 37.23 \\
6 & 63.55 & \underline{48.57} & \underline{44.45}
  & 59.19 & \underline{44.31} & \underline{39.41}
  & \underline{56.52} & \underline{42.28} & \underline{37.41} \\
8 & \textbf{65.18} & \textbf{50.00} & \textbf{45.80}
  & \textbf{60.66} & \textbf{45.92} & \textbf{40.83}
  & \textbf{57.86} & \textbf{43.73} & \textbf{38.77} \\
\bottomrule
\end{tabular}
\end{table}

\begin{table}[h]
\centering
\caption{Parameter sensitivity of ADS-POI w.r.t. the embedding dimension $d$ under the single-state setting ($K=1$). Best results are in \textbf{bold}, and second-best results are \underline{underlined}.}
\label{tab:param_sensitivity_d}
\small
\begin{tabular}{c|ccc|ccc|ccc}
\toprule
\multirow{2}{*}{$d$}
& \multicolumn{3}{c|}{\textbf{NYC}}
& \multicolumn{3}{c|}{\textbf{TKY}}
& \multicolumn{3}{c}{\textbf{CA}} \\
\cmidrule(lr){2-4} \cmidrule(lr){5-7} \cmidrule(lr){8-10}
& \textbf{HR@10} & \textbf{NDCG@10} & \textbf{MRR}
& \textbf{HR@10} & \textbf{NDCG@10} & \textbf{MRR}
& \textbf{HR@10} & \textbf{NDCG@10} & \textbf{MRR} \\
\midrule
32
& 54.60 & 38.87 & 34.68
& 49.92 & 35.41 & 31.28
& 47.38 & 33.76 & 29.64 \\
64
& \underline{55.34} & \underline{39.56} & \underline{35.22}
& \underline{50.71} & \underline{36.18} & \underline{31.94}
& \underline{48.12} & \underline{34.47} & \underline{30.31} \\
128
& \textbf{55.81} & \textbf{40.02} & \textbf{35.68}
& \textbf{51.24} & \textbf{36.74} & \textbf{32.41}
& \textbf{48.67} & \textbf{35.03} & \textbf{30.88} \\
256
& 55.62 & 39.84 & 35.51
& 51.06 & 36.52 & 32.18
& 48.53 & 34.81 & 30.63 \\
\bottomrule
\end{tabular}
\end{table}

\end{document}